\documentclass[aps,prd,showpacs,preprint,amsfonts]{revtex4-1}
\usepackage{epsfig}
\usepackage{graphics}
\usepackage{amsmath}
\newcommand{\Eq}[1]{Eq.~(\ref{#1})}
\newcommand{\Eqs}[1]{Eqs.~(\ref{#1})}

\newcommand{\be}{\begin{equation}}
\newcommand{\ee}{\end{equation}}
\newcommand{\lambdat}{\mathchar'26\mkern-9mu \lambda_t}
\newcommand{\lambdap}{\mathchar'26\mkern-9mu \lambda_p}
\newcommand{\bea}{\begin{eqnarray}}
\newcommand{\eea}{\end{eqnarray}}
\def\half{{\scriptstyle{\frac{1}{2}}}}
\def\fourth{{\scriptstyle{\frac{1}{4}}}}
\def\threehalf{{\scriptstyle{\frac{3}{2}}}}

\def\threefourth{{\scriptstyle{\frac{3}{4}}}}

\begin{document}

\author{Jacob D. Bekenstein}
\affiliation{Racah Institute of Physics, Hebrew University of
Jerusalem, Jerusalem 91904, Israel\\}\date{\today}
\title{Optimizing entropy bounds for macroscopic systems}
\begin{abstract}
The universal bound on specific entropy (entropy to energy ratio) was originally inferred from black hole thermodynamics.   We here show from classical thermodynamics alone that for a system at fixed volume or fixed pressure, the entropy to energy ratio $S/E$ has a unique maximum, $(S/E)_\mathrm{max}$.  A simple argument from quantum dynamics allows one to set a model--independent upper bound on $(S/E)_\mathrm{max}$ which is usually much tighter than the universal bound.  We illustrate with two examples.

\end{abstract}
\maketitle

\section{Introduction}
\label{sec:intro}
Contemporary theoretical physics puts great stock on the holographic principle and its associated mathematical machinery.  That principle promises to  reduce complex dynamical problems in high energy and condensed matter physics to the solution of simpler problems in gravitational theory.  The holographic principle was inspired by the holographic entropy bound~\cite{thooft,susskind}, which derived from black hole physics just as did the earlier universal entropy bound~\cite{bek81}.  This last named bound places a much stronger upper bound on the specific entropy (entropy per unit energy) of a system than can the holographic bound~\cite{bekSciAm}, but not strong enough to be of practical use, for example in limiting the information capacity of a a planned state-of-the-art computer memory.  

The universal entropy bound first arose from a \textit{gedanken} experiment in which one drops an ordinary system into a black hole, and requires that the generalized second law of thermodynamics~\cite{bek74} be upheld.  Despite gravity's involvement in the deduction, the resulting bound on entropy $S$,
\be
\frac{S}{E}<\frac{2\pi R}{\hbar c}\,,
\label{unbound}
\ee
where $R$ stands for the circumscribing radius of the system and $E$ for its total energy,
does not involve $G$.  The bound was thus interpreted~\cite{bek81} as applying principally to ordinary systems with weak gravitational fields.  Indeed, for systems generating strong gravitational fields, such as matter already collapsing inside a black hole, bound~(\ref{unbound})  can fail~\cite{bousso}.

Concurrent with the inference of the universal entropy bound from gravitational physics, it was shown from quantum statistical mechanics that any ordinary system in equilibrium has a maximum specific entropy (entropy per unit energy)  which is reached at the temperature at which the free energy of the system in question vanishes~\cite{bek81}.  For some model systems one can explicitly calculate the maximum specific entropy, and this is found to accord with the universal entropy bound~\cite{bek84}. 

In the universal entropy bound the energy refers to all the energy, including rest energies.  Indeed the appearance of ``$c$" in the bound reminds us that singling out the nonrelativistic energy is not allowed.  But in many applications ranging from condensed matter physics to chemistry, the rest masses are not customarily included in the energy of a system whose entropy is of interest.  Is there then some equally strict bound on the ratio of entropy to nonrelativistic energy?   Such bound would bring us closer to the ideal of usefully constraining the entropy of a generic statistical system or the information capacity of a generic memory. 

In this paper we explore the existence of tighter entropy bounds for macroscopic systems.  In Sec.~\ref{sec:vol} we show that a thermodynamic system at fixed volume (or other extensive variable) exhibits a unique maximum of the ratio of entropy to thermal energies.  This is true even in the face of phase transitions.  Sec.~\ref{sec:press} extends this result to systems at fixed pressure (or other intensive variable).  The above maxima are each given by a certain characteristic reciprocal temperature, $1/T_0$ and $1/T_1$, respectively.  In Sec.~\ref{sec:T-tau} we rework  an argument by Landau and Lifshitz that sets an upper bound on the accessible reciprocal temperature of a generic macroscopic system.  The resulting bound serves as an upper bound on the ratio of entropy to energy.  In Sec.~\ref{sec:macroex} we give two examples of the consequences of the new entropy bound, from which it is clear that it is much tighter than the universal one.  Sec.~\ref{sec:sum} summarizes our findings and explores possible future work on extending the bounds to microscopic systems.

Henceforth we adopt units in which Boltzmann's constant $k_\mathrm{B}$ is unity.

\section{Peak $S/E$ at fixed volume}
\label{sec:vol}

Consider a thermodynamic system in equilibrium containing a fixed amount of matter; assume an extensive variable in it is held constant.  For concreteness we take the fixed variable to be the volume $V$, but the arguments is easily extended to other types of systems.  In our example the other thermodynamic variables are the entropy $S$, the total energy $E$, the temperature $T$ and the pressure $P$.  From the expression for the first law,
\be
T dS = dE + P dV = dE
\label{firstlaw1}
\ee
we have that
\be
\frac{1}{T}=\left(\frac{\partial S}{\partial E}\right)_V.
\label{T}
\ee 
Differentiating with respect to $E$ at constant $V$ gives
\be
\left(\frac{\partial^2 S}{\partial E^2}\right)_V=-\frac{1}{T^2}\frac{1}{(\partial E/\partial T)_V}\ .
\label{spheat}
\ee

\begin{figure}
\includegraphics[width=3.0in]{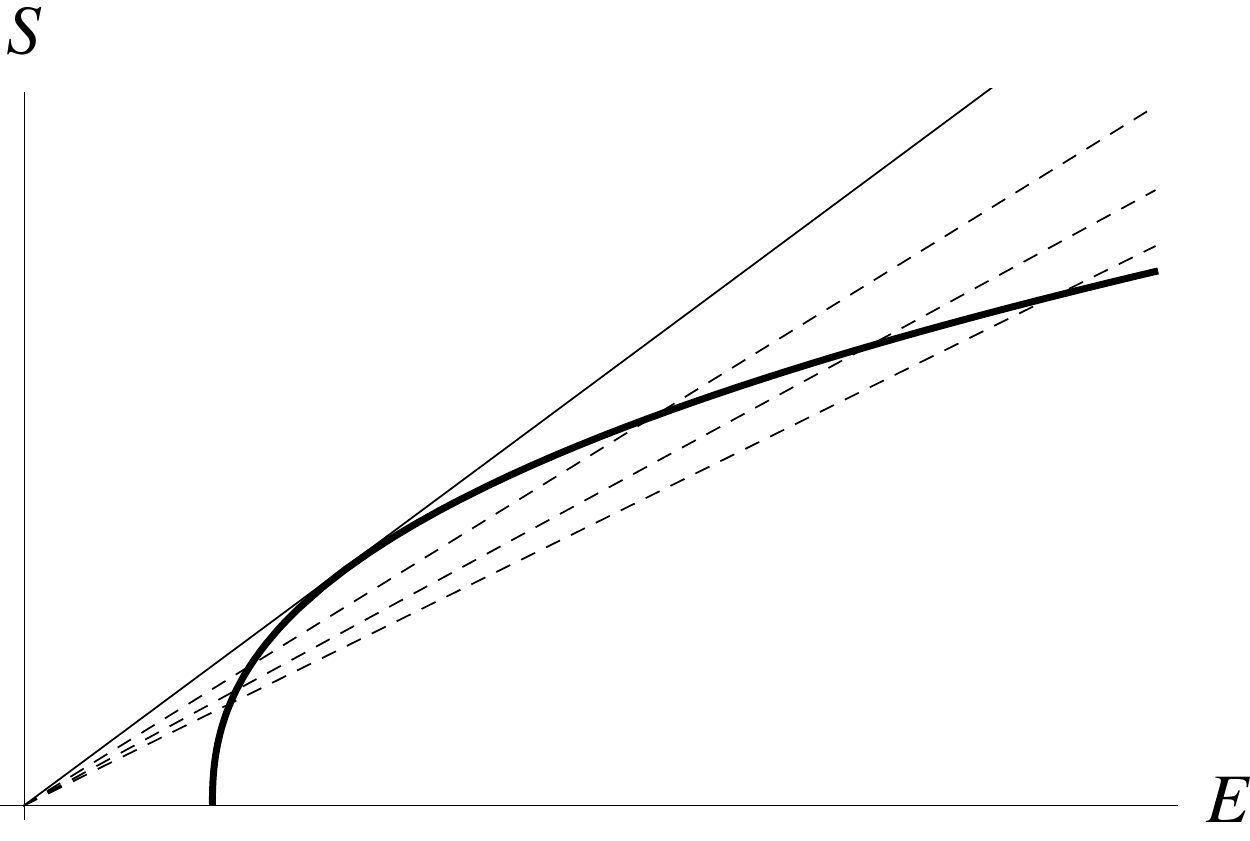}
\caption{Schematic $S(E)$ curve (solid).  The slope of any straight line through the origin that intersects the curve, e.g. dashed lines, gives the $S/E$ at the $E$ of intersection.   It may be seen that the maximum $S/E$ equals the slope of the tangent to the curve (thin solid line).\label{Fig1}}
\end{figure}

\begin{figure}
\includegraphics[width=3.0in]{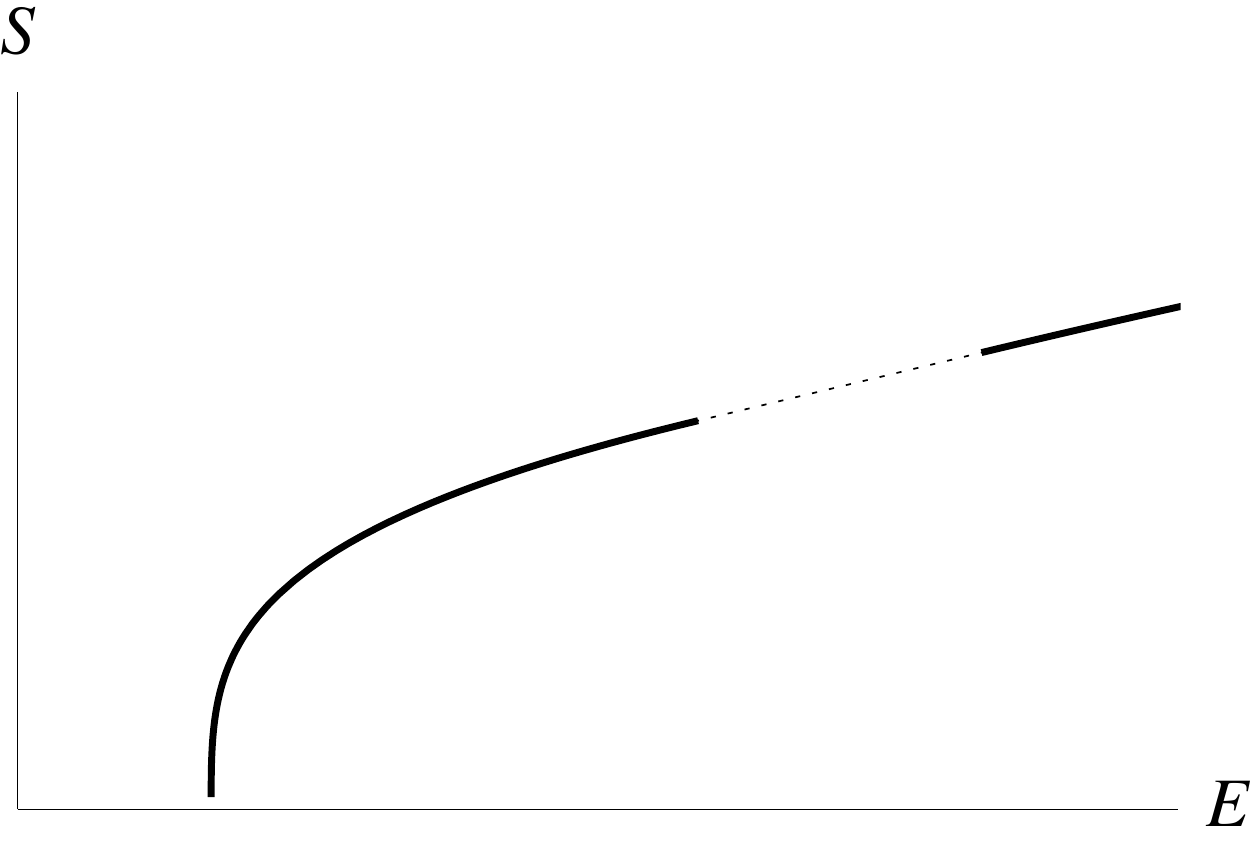}
\caption{The $S(E)$ curve for a system that undergoes a first order phase transition.  The straight dotted line segment is the locus of states of the system in the course of the transition.  It may be seen that there is a unique tangent to the  curve that also passes through the origin.  In the situation pictured the maximum $S/E$ is attained below the transition, but by properly shifting the foot of the curve we can arrange for it to be reached during the phase transition, or above it.
\label{Fig2}}
\end{figure}

Because the system is presumed to be in a canonical ensemble, the heat capacity at constant volume, ${(\partial E/\partial T)_V}$, must be positive definite.   It follows that $\left({\partial^2 S}/{\partial E^2}\right)_V$ is nonpositive.  In most cases this means that the curve of $S(E)$ is concave (downward) everywhere.  The exception is a first-order phase transition, during which the heat capacity diverges, so that the curve $S(E)$ is straight throughout the course of it (see Fig.~\ref{Fig2}).  We consider $E$ to always include some component besides thermal energy.  Thus $S(E)$ passes through zero at some positive $E$; in the absence of the phase transition its curve looks qualitatively as illustrated in Fig.~\ref{Fig1}.  By \Eq{T}  the curve's slope must diverge at $S=0$ ($T=0$ at $S=0$).

Focus on Fig.~\ref{Fig1}. Geometrically the slope of a straight line through the origin that intersects the curve gives the value of $S/E$ at the intersection(s).  It is immediately evident that the particular straight line through the origin  which is tangent to the curve of $S(E)$ has a slope equal to $(S/E)_\mathrm{max}$, the peak value taken on by $S/E$.  It is also obvious that the (local) maximum in question is unique.  At the point of tangency the slope of the $S(E)$ curve is $(\partial S/\partial E)_{\rm V}$ which by \Eq{T} amounts to $1/T_0$, where $T_0$ is the temperature at the point where $S/E$ is maximal.  But obviously the two slopes are equal so that
\be
\left(\frac{S}{E}\right)_\mathrm{max} =\frac{1}{T_0}\,.
\label{bound1}
\ee
Of course $T_0$ is a function of the fixed volume assumed.  From the definition of the Helmholtz free energy, $F\equiv E-TS$, it follows immediately that $F(V,T=T_0)=0$.  Both of these conclusions agree with the inferences drawn long ago from statistical mechanics~\cite{bek81}. 

The requirement that $S(E)$ vanish at a positive $E$ means in some cases that one has to shift the natural entropy vs energy curve to move its foot (the $S=0$ point) away from the origin.  A case in point is pure thermal radiation of massless quanta for which $S$ at fixed $V$ behaves as the $3/4$ power of the thermal energy.  The question arises, how fast does $(S/E)_\mathrm{max}$ shift as the zero of $E$ is so shifted?  Let us write $E=\tilde E+\epsilon$, where $\tilde E$ is the purely thermal energy while $\epsilon$ is the shift of the foot of the entropy curve from the origin in Fig.~\ref{Fig1}.  Obviously the entropy itself is given by some $\epsilon$ independent function $\mathcal{S}(\tilde E)$ such that $\mathcal{S}(\tilde E)=S(\tilde E+\epsilon)$ and $\mathcal{S}(0)=0$.   Our question may now be rephrased: what is the change \emph{of the maximum} of $\mathcal{S}(\tilde E)/(\tilde E+\epsilon)$ under the infinitesimal shift $\epsilon\to \epsilon+\delta\epsilon$?    We might venture the guess  that since $S/E$ has dimensions of reciprocal energy, the shift should occur at the same rate as that of $1/\epsilon$.  As we show now, $S/E$ is less sensitive to the shift than this guess would lead us to believe.

Obviously the value of $\tilde E$ at the maximum of $S/E$ shifts as $\epsilon$ changes.  We suppose henceforth that $\delta\epsilon>0$.   Writing $\mathcal{S}'$ for $d\mathcal{S}(\tilde E)/d\tilde E$ we have
\be
\delta\, \frac{\mathcal{S}}{\tilde E+\epsilon}=\left(\mathcal{S}'-\frac{\mathcal{S}}{\tilde E+\epsilon}\right)\frac{\delta\tilde E}{\tilde E+\epsilon}-\frac{\mathcal{S}\,\delta\epsilon}{(\tilde E+\epsilon)^2}=-\frac{\mathcal{S}\,\delta\epsilon}{(\tilde E+\epsilon)^2}\,.
\ee
The term in round brackets has dropped out because $S/E$ at its maximum equals the reciprocal temperature $\mathcal{S}'$ there.  The result can be cast as
\be
\delta \ln\frac{\mathcal{S}}{\tilde E+\epsilon}=-\frac{\delta\epsilon}{\tilde E+\epsilon}>-\delta\ln \epsilon\,,
\ee
where the inequality comes about because $\tilde E>0$.  The result tells us that $\epsilon S/E$ is a growing function of $\epsilon$.  This is a nontrivial finding since a moment's inspection of Fig.~\ref{Fig1} will reveal that $S/E$ decreases with increasing $\epsilon$.  Therefore, regardless of the precise shape of $\mathcal{S}(\tilde E)$, $(S/E)_\mathrm{max}$ decreases with increasing $\epsilon$ more slowly than does $1/\epsilon$.  Thus for some shapes of $\mathcal{S}$ the peak value of $S/E$  may be insensitive to a shift of the foot of the curve.

Thus far we implicitly assumed that $S(E)$ in an analytic function.  In the presence of a phase transition of the first order this is no longer  true.  An example would be a sealed flask containing some water ice.  As we slowly raise the temperature there comes a certain temperature $T_p$ at which the ice begins to melt.  We know this entails a growth of the system's entropy because the molecules in the liquid are less ordered than those in the solid.  According to \Eq{firstlaw1} the system's energy grows (at the expense of the thermal bath which defines the temperature).  We know that the whole melting of the ice takes place at fixed temperature, $T=T_p$, so that over the entire process we must have $\Delta E=T_p\,\Delta S$. Thus in the $E-S$ diagram the phase transition is a segment of a straight line with slope $1/T_p$ bridging the two segments of the curve $S(E)$ which lie above and below the transition, as illustrated in Fig.~\ref{Fig2}.

Of course the ratio $\Delta E/\Delta S=T_p$ equals the temperature both at the melting's inception  and at its end.  Therefore, $1/T_p$  coincides with the slopes of the $S(E)$ curve just before the phase transition begins and just after its end.  Hence the slope of the straight segment coincides with the slope of the $S(E)$ curve just before and just after the transition.  It follows that the locus of $S(E)$ is continuous and smooth, as illustrated in Fig.~\ref{Fig2}. 

\section{Peak $S/E$ at fixed pressure}
\label{sec:press} 

How do the previous conclusions change when the system is one held a fixed pressure $P$ (or fixed intensive variable)?  To answer this we shall focus on the enthalpy $W\equiv E+PV$.  Evidently the first law takes the new form
\be
T dS = dW - V dP = dW
\label{firstlaw2}
\ee
so that
\be
\frac{1}{T}=\left(\frac{\partial S}{\partial W}\right)_P.
\label{T1}
\ee
In analogy with \Eq{spheat} we obtain
\be
\left(\frac{\partial^2 S}{\partial W^2}\right)_V=-\frac{1}{T^2}\frac{1}{(\partial W/\partial T)_P}\ .
\ee 
Of course the heat capacity at constant pressure, ${(\partial W/\partial T)_P}$ must exceed the one at constant volume (since the system does work on its surroundings), and so is also positive.
It follows, as in Sec.~\ref{sec:vol}, that the curve of $S(W)$ is concave (downward) everywhere.  With the replacement $E\to W$, Fig.~\ref{Fig1} may also represent the generic $S(W)$.

As before we arrange that the curve has $S=0$ at a positive $W$.  Then the geometric argument in Sec.~\ref{sec:vol} may be repeated for the $W-S$ plane with the conclusion that
\be
\left(\frac{S}{W}\right)_\mathrm{max} =\frac{1}{T_1}\ ,
\label{bound2}
\ee
where $T_1$ is the temperature at which $S/W$ attains its unique local and global maximum.  From the definition of Gibbs free energy, $G\equiv W-TS$, it follows that $G(P, T=T_1)=0$.

Further, one may allow a phase transition at constant pressure. According to \Eq{firstlaw2} and the mentioned principle that entropy jumps at a phase transition we may repeat the argument given in Sec.~\ref{sec:vol} to conclude that the phase transition is represented by a straight line segment continuously and smoothly bridging the two curved parts of $S(W)$, just as shown in a suitably relabeled version of Fig.~\ref{Fig2}.

Our conclusion at the end of Sec.~\ref{sec:vol}, that with increasing $\epsilon$, $(S/E)_\mathrm{max}$ decreases more slowly than $1/\epsilon$, is valid also in the context of a system held at a constant intensive variable.

\section{Semiclassical upper bound on $S/E$}
\label{sec:T-tau}

In practice, to set a bound on $S/E$ or $S/W$ we must know $T_0$ or $T_1$.  Of course each is determined as the zero of $F(T)$ or $G(T)$, respectively. However, we would gain little by appealing to this fact; it would be as easy to calculate $(S/E)_\mathrm{max}$ or $(S/W)_\mathrm{max}$ directly from the $S, E$ or $W$ derived thermodynamically from $F(T)$ or $G(T)$. What we really need is some \emph{independent lower bound} on $T_0$ or $T_1$.  

Landau and Lifshitz's (henceforth LL) have given a lower bound on the temperature of a macroscopic system~\cite{LL} expressed in terms of the timescale $\tau$ for change of a well defined system observable $x$.  They represent $x$ by an Hermitian operator $\hat x$.  Assuming $\hat x$ it is not a conserved quantity, they focus on the \emph{semiclassical} Heisenberg equation of motion  
\be
[\hat x,\hat H]=\imath\hbar\, \dot x\approx \imath\hbar\, x/\tau,
\label{semi}
\ee
where $\hat H$ is the system's quantum Hamiltonian and $\dot x$ on the r.h.s. is a c-number, the Poisson bracket of the classical $x$ and $H$.  \Eq{semi} is viewed as a commutation relation, and from it  follows the uncertainty relation~\cite{Merzbacher}   $\Delta x\Delta H\geq \half \hbar |x|/\tau$.  By asserting that the system as a whole is far from  the quantum regime, LL justify assuming that $\Delta x\ll |x|$, so that 
\be
\Delta H\gg \hbar/\tau\,.
\label{uncert}  
\ee

In the second part of the argument, LL regard the entropy $S$ of the system as a function of its momentary energy $\varepsilon$, and focus on $\exp[S(\varepsilon)]$ as proportional to the probability of the state with energy $\varepsilon$.  In order that this probability be stable against quantum fluctuations, LL require that $\Delta H< T$ (recall that $dS/d\varepsilon)=1/T$).  Thus for a macroscopic system (for which quantum fluctuations are negligible) LL obtain the bound
\be
T\gg \hbar/\tau\qquad (\mathrm{Landau\ and\ Lifshitz}).
\label{Tbound}
\ee

This second part of the argument is obscure.  In canonical ensemble temperature is defined as the energy derivative of the entropy of the \emph{reservoir} with which the system is in equilibrium. But in LL's argument the derivative is taken of the system's energy. Further,  the conclusion is counterintuitive.  We should expect that a large system---with many degrees of freedom---have a bigger $\Delta H$ than a small one,  if both are at the same temperature. This effect is not apparent in LL's requirement that $\Delta H<T$.  In view of all these we replace the second part of LL's argument by the following.

We appeal to the elementary statistical mechanics of a system in canonical ensemble, that is at a given temperature $T=1/\beta$.  The function to focus on is the partition function $Z(\beta)$, which is derivable from the Hamiltonian.  We know~\cite{Reif} that  $E=-\partial \ln Z(\beta)/\partial\beta$ as well as $\Delta H^2=\partial^2 \ln Z(\beta)/\partial\beta^2$.  It follows that
\be
\Delta H^2=-\partial E/\partial \beta=T^2\partial E/\partial T=T^2 C,
\label{DH2}
\ee 
where $C$ is the system's appropriate type of heat capacity---not the specific heat.  This is a dimensionless number because we take $k_\mathrm{B}=1$. Thus in contrast with LL's requirement $\Delta H<T$, we have here the result
\be
\Delta H = \surd C\, T.
\label{DH}
\ee
This does display the expected growth of the energy dispersions with the size of the system.  Together with \Eq{uncert} it gives, in lieu of \Eq{Tbound}, the new lower bound on the temperature of a macroscopic system,
\be
T\gg \frac{\hbar}{\surd C\, \tau}\,.
\label{T>}
\ee
We stress, as would LL, that this does not preclude existence of lower temperatures; it only claims that a lower temperature can occur only in a system that, because of quantum fluctuations, cannot be treated by ordinary thermodynamics of macroscopic systems.

Combining \Eqs{bound1}, (\ref{bound2}) and (\ref{T>}) we get
\be
\left(\frac{S}{E}\right)_\mathrm{max}\ll \frac{\surd C_V\, \tau}{\hbar};\qquad \left(\frac{S}{W}\right)_\mathrm{max}\ll\frac{\surd C_P\, \tau}{\hbar}\,. 
\label{newbounds}
\ee
Again, we mean that a macroscopic system treatable by ordinary thermodynamics cannot have arbitrarily large entropy-to-energy ratios, as shown.

When applying these inequalities, one should not be required to compute $\tau$ \emph{ab initio}.  After all the idea behind the entropy bound is to supply a quick and easy estimate of the maximum entropy given the energy.  A physical estimate of $\tau$ should suffice.  This $\tau$ can turn out to be temperature dependent.  For example, in a classical gas the time for changes should be longer at lower temperatures when the molecules travel slower.  Likewise, a $C$ can change at the temperature at which a class of degrees of freedom is frozen out.  And, of course, by considering different variables $\hat x$, we can come up with different $\tau$s.  Thus in applying \Eqs{newbounds} one should attempt to reduce the r.h.s. to a minimum by taking cognizance of all these options.  Such procedure will simultaneously optimize the lower bound on temperature (\ref{T>}). 

As will be clear from the examples below, the ``much smaller than'' symbol ``$\ll$" in \Eq{newbounds} can often be traded for an ``approximately equal or smaller than'' symbol ``$\lesssim$", in the process of which the $C$ disappears from the inequality.  The new bounds, although less precisely stated than the universal entropy bound, can be shown to be more stringent than it.  The universal bound refers to the system's total energy, including rest energy, while the present bounds invoke only the thermal energy (and a small amount of some other energy as explained in Sec.~\ref{sec:vol}).  The ratio of the two energies is of order $c^2/v^2$ where $v$ is the typical particle velocity in the system.  On the other hand, the time scale $\tau$ mentioned in the bounds~(\ref{newbounds}) is typically longer than the light crossing time $R/c$ invoked by the universal bound by a factor equaling the ratio of $c$ to the sound speed (or what is often similar, the particle velocity $v$).  Thus bounds~(\ref{newbounds}) are tighter than bound~(\ref{unbound}) by a factor of order $c/v\gg 1$.
\section{Examples}
\label{sec:macroex}

\subsection{Monatomic Boltzmann gas}
\label{sec:Boltzmann}

Consider a nonrelativistic monatomic gas consisting of $N$ atoms confined to a cubical box of size $R$.  At temperature $T$ the typical atomic velocity in one direction is $\sim \sqrt{2T/m}$.  Therefore the time between successive collisions with the walls is $\sim  R\sqrt{m/2T}$.  This can be taken as $\tau$.  Alternatively, we can think of a disturbance of the pressure in the form of a sound wave.  This crosses the box in a time $\tau\approx R/c_s$ where $c_s$ is the sound speed at the temperature in question. However, $c_s$ is of the order of the atomic velocity, so we end up with the same $\tau$ as before.

It might be objected that $\tau$ should rather be taken as the time between interatomic collisions.  It is, however, unclear whether this timescale is connected with time variation of a global system variable.  In any case we could assume that the density is so low (a Knudsen gas) that the mean free path is longer than $R$.  Thus we can take $\tau\approx  R\sqrt{m/2T}$. 

With regard to the gas' energy we must remember that the present approach requires the zero of energy to be set so that the energy remains positive as $S$ vanishes.  Therefore we shall take $E$ to be the sum of the thermal energy and the quantum zero point energy.  This latter is  $\approx N(\hbar/R)^2/2m$; in assuming that the gas is a Boltzmann gas we take it that all atoms can sink to the lowest available translation energy level.  The thermal energy is $\threehalf NT$ (for temperatures not so low that the gas behaves classically).   We also learn from this that $C=\threehalf N$.   Substituting all this in the first case of \Eq{newbounds}  and dividing through by $\half\sqrt{m/T}$ we have
\be
\frac{S/N}{\fourth\hbar^2/R^2\sqrt{1/mT}+\threefourth\sqrt{mT}}\ll \frac{\sqrt{3N} R}{\hbar}
\ee

We could leave this as a strong inequality; however, we can improve on it.  Recall  that the  ``$\ll$'' comes in because we suppose that in a macroscopic system $\Delta x\ll \langle x\rangle$.  For the $N$-body system the relative uncertainty of a collective quantity will certainly scale as $1/\surd N$.  Thus we may replace ``$\ll$'' above by $ \lesssim1/\surd N$ obtaining
\be
\frac{S/N}{\fourth(\hbar/R)^2\sqrt{1/mT}+\threefourth\sqrt{mT}}\lesssim \frac{\sqrt{3N} R}{\surd N\hbar}=\frac{\surd 3 R}{\hbar}\,.
\ee

Of course we do not contemplate dealing with a quantum gas.  Thus $R$ should be at least a few times the reduced thermal de Broglie wavelength $\lambdat\equiv \hbar/\sqrt{3 m T}$.  This requirement makes the second term of the sum in the denominator dominate over the first.  Thus 
\be
S/N \lesssim 3R/4\lambdat.
\label{S/N}
\ee
By contrast, according to the exact expression for the entropy of an ideal monoatomic gas~\cite{LL}, $S/N$ is of order unity for $R \sim \lambdat$ and rises logarithmically with $R/\lambdat$ at larger $R$.  This behavior is perfectly compatible with bound~(\ref{S/N}) and therefore with the new entropy bound~(\ref{newbounds}).

\subsection{Dielectric solid at low temperature}
\label{sec:dielectric}

Consider a solid block of dielectric material whose three linear dimensions are similar and whose volume is $V$.  In a dielectric the specific heat at low temperatures is due to phonons. Well below the Debye temperature the number and energy of the phonons can be approximated by~\cite{LL}
\be
N_\mathrm{ph}=\frac{2\zeta(3)\, T^3 V}{\pi^2\hbar^3 \bar c_s{}^3};\qquad  E_\mathrm{ph}=\frac{\pi^2 T^4 V}{15 \hbar^3 \bar c_s{}^3}\,.
\label{E&N}
\ee
Here $\bar c_s{}^3$ is the harmonic mean of the cubes of the longitudinal and the two transversal sound speeds ($3/\bar c_s{}^{3}=1/c_l{}^{3}+2/c_t{}^{3} $).  From \Eq{E&N} it follows that
\be
C_V=\frac{4\pi^2 T^3 V}{15 \hbar^3 \bar c_s{}^3}\,=\frac{ 2\,\pi^4 N_\mathrm{ph}}{15\,\zeta(3)}.
\ee

Viewed as sound waves the phonons should have associated with them some collective variable varying on time scale $V^{1/3}/\bar c_s$.  Of course we should also find phonon frequencies higher than $\bar c_s/V^{1/3}$, but these would not naturally correspond to a global motion of the material.  Hence we take $\tau\approx V^{1/3}/\bar c_s$.

We now substitute all the above into \Eq{newbounds}; as done in Sec.~\ref{sec:Boltzmann} and for like reason we shall replace ``$\ll$''  by $ \lesssim1/\surd N_\mathrm{ph}$: 
\be
\left(\frac{S}{E}\right)_\mathrm{max}\lesssim \frac{3.287\, V^{1/3}}{\hbar \bar c_s}\ .
\label{newbounds1}
\ee

Let us contrast this with bound (\ref{unbound}).  Of course $V^{1/3}$ here corresponds closely to the $R$ in bound (\ref{unbound}).  But here $\bar c_s$---the speed of sound in a solid---is a factor $10^5$ lower than $c$ in (\ref{unbound}).  By contrast in  the universal bound 
 $E$ must include all forms of energy in the system while in \Eq{newbounds1} $E$ need only include just enough of a nonthermal variety of energy to keep it from vanishing as $S\to 0$.  Now in a solid \emph{at room temperature} the thermal energy is about $\threehalf T$ per atom or about $5\times 10^{-14}$ ergs; the thermal energy would be much smaller at low temperatures.  By contrast the rest energy per atom is about $0.02$ erg.  Thus in the present problem the optimized bound  \emph{on entropy} is tighter than what the the universal bound (\ref{unbound}) would give by at least a few times the factor  $0.02/5\times 10^{-14}/10^5\approx 4\times 10^7$.

It is also possible to employ bound (\ref{newbounds1}) in reverse; it can be viewed as putting a lower bound on the linear dimension $V^{1/3}$ of the chunk of dielectric being contemplated.  Once we have included in $E$ the suitable extra kind of energy to render the peak $S/E$ finite, the temperature $T_0$ is defined.  To it corresponds a phonon Planckian frequency  distribution which peaks at $\omega=\omega_p\equiv  2.822 T_0/\hbar$~\cite{LL}.  The reduced wavelength of phonons at the peak would be $\lambdap=\bar c_s/\omega_p$.  From \Eq{newbounds1} it thus follows that
\be
V^{1/3}  \gtrsim 0.86 \lambdap.
\ee
This is reasonable.  Treatment of phonons as a thermal gas only makes sense in a space which is larger than the typical wavelength of the majority of phonons.  Were we to employ bound~(\ref{unbound}) for the same purpose it would set a lower bound on the linear size $R$ which is a factor $c/\bar c_s$ larger than that we have just obtained; such bound would certainly be overly cautious.

\section{Summary and outlook}
\label{sec:sum}

We have shown that for every non-relativistic thermodynamic system in equilibrium the ratio of entropy to energy (not including rest masses) has a unique local maximum, regardless of whether the system is held at fixed volume or at fixed pressure, or whether it passes through phase transitions.  The argument is purely thermodynamical.  By a further modest appeal to quantum theory we are able to set an upper bound on the said ratio in terms of a certain timescale and of the heat capacity of the system.  We also display a more informal but more transparent form of this bound, which, although less precise than the universal entropy bound, is many orders of magnitude tighter than it. 

The appeal we made to semiclassical concepts in Sec.~\ref{sec:T-tau} limits the use of the bounds given here to macroscopic systems.  But there are reasons for desiring like bounds applicable to microscopic systems.  With the rise of nanoscience there has developed the need to understand the thermodynamics of systems made of a small number of atoms.
And with the advances in computing and information systems has come the need to deal with memories of microscopic dimensions.   Both of these directions are closely concerned with the entropy \emph{capacity} of microscopic systems; hence the need for tight entropy bounds to make the easy assessment of the capabilities of generic quantum systems possible.

The line of reasoning espoused here can be of help also in the microscopic realm.  But first the notion of the timescale $\tau$ required in those arguments has to be clarified in this connection.  One impediment is that, in a strict sense, a thermal system, by being stationary, has no finite timescale associated with it.  This impediment has been sidestepped in the macroscopic case by taking a semiclassical point of view, \emph{a la} Landau and Lifshitz.  This is no longer possible when treating a microscopic system entirely within quantum statistical physics.  The consequent need to consider a perturbed thermal system considerably complicates the arguments.

\acknowledgments

I thank David E. Bruschi for help and Ofer Lahav and Jonathan Oppenheim for their hospitality at University College London where this paper was put in final shape.   This research is supported by the I-CORE Program of the Planning and Budgeting Committee and the Israel Science Foundation (grant No. 1937/12), as well as by the Israel Science Foundation personal grant No. 24/12.

\end{document}